\documentclass[twocolumn,showpacs,preprintnumbers,amsmath,amssymb]{revtex4}
\usepackage{mathrsfs}
\usepackage{txfonts}
\usepackage{graphicx}% Include figure files
\usepackage{dcolumn}% Align table columns on decimal point
\usepackage{bm}% bold math
\usepackage{slashed}% \slashed{A}~a\!\!\!/
\usepackage{color} % \textcolor[rgb]{1.00,1.00,0.50}{YourText} or {\color{blue}YourText}
 {\catcode`\|=\active \gdef|{\egroup\,\vrule\,\bgroup}}

\def\be{\begin{eqnarray}}
\def\ee{\end{eqnarray}}
\begin{document}

%\begin{CJK*}{GBK}{song}
%\preprint{PRA/Version 1}

\title{Optimal attosecond pulses generation from oscillating plasma surfaces driven at relativistic intensity}% Force line breaks with \\

%
%\author{Yue Guo, Aihua Liu\email{aihualiu@jlu.edu.cn}, Jun Wang\email{wangjun86@jlu.edu.cn} and  Xueshen Liu}\email{aihualiu@jlu.edu.cn, wangjun86@jlu.edu.cn}
%\author{Yue Guo, Aihua Liu}\email{aihualiu@jlu.edu.cn} \author{Jun Wang}\email{wangjun86@jlu.edu.cn} \author{Xueshen Liu}
\author{Lin Cheng$^{1,2*}$, Zhixin Fan$^{1,3}$} \thanks{These authors contributed equally.} 
\author{Liguang Jiao$^3$ }\email{lgjiao@jlu.edu.cn}
\author{Aihua Liu$^1$ }\email{aihualiu@jlu.edu.cn}
\affiliation{1. Institute of Atomic and Molecular Physics, Jilin University, Changchun 130012, China}
\affiliation{2. School of Physics, Peking University, Beijing 100871, China}
\affiliation{3. College of Physics, Jilin University, Changchun 130012, China}
%\affiliation{Jilin Provincial Key Laboratory of Applied Atomic and Molecular Spectroscopy, Jilin University, Changchun 130012, China}

%\affiliation{

\date{\today}% It is always \today, today,

\begin{abstract}
{The creation of attosecond pulses via laser-plasma interaction has been a subject of great scientific interest for more than three decades. 
This process is investigated by using particle-in-cell simulation with varying the plasma and laser parameters. The steepness of the density gradient at the plasma-vacuum interface is examined to see how this parameter affects the high-order harmonic generations and isolated-attosecond pulse creation. The optimal density gradient lengths $L$ are explored within the relativistic oscillating mirror mechanism. 
Although the ideal gradient lengths for the full width at half maximum and peak intensity of an isolated-attosecond pulse depend on the driving laser intensity independently, they are both found near $L$=0.2$\lambda$ for high laser intensities.

 }
\end{abstract} \vskip 0.5in

\maketitle

\vskip 1in%

   \section{Introduction}

A new era in science has begun with the invention of the laser. Since the advent of the first laser pulse six decades ago, the laser has been widely used in a variety of science and industries \cite{Pertot17,Hu06,Hass97,Dobo05,Lara18,Cale14,Taka15,Krau16,Chini14,San11,Krau09,Zhu2021}. Scientists are constantly searching for new ways to create lasers with higher peak intensity and shorter pulse duration. The chirped pulse amplification technique developed by D. Strickland and G. Mourou \cite{Strickland85} allows the laser pulse to be amplified to a peak intensity high enough to be applied to the laser-plasma interaction. On the other hand, an ultrashort laser pulse can now be easily produced in laboratories, the pulse duration has been drastically reduced from nanosecond (1 ns = $10^{-9}$ s) in the 1960s to attosecond (1 as = $10^{-18}$ s) today \cite{Li17, Gaum17}.

In 2001, the first attosecond pulse, a light pulse with the pulse duration of less than 1 femtosecond (1 fs = $10^{-15}$ s), was produced by M. Hentschel {\it et al.} \cite{Krau01}. Since the attosecond pulse has a high time resolution and wide spectrum that covers the water window, it can be used to study ultrafast processes and gain insight into subatomic processes \cite{Li17, Gaum17}. Using such a short laser pulse, one can observe and control the internal dynamics of atoms and molecules \cite{Krau09, Krau16}. This technology could indeed be used to detect the electronic dynamics of an atomic electron shell.

%At present, most of the studies of attosecond pulses focus on the production of high order harmonics generation (HHG) from gases, solids and liquids targets, and then filtering to obtain the attosecond pulses. Studies have shown that the attosecond pulses intensity or conversion efficiency in these ways are still at low level. The advantage in using plasma harmonics instead is by the orders of magnitude higher expected intensity of the attosecond pulses. This will not only greatly ease their characterization, but also significantly expand the scope of their applications.

%\begin{figure}
%\includegraphics[width=1.0\columnwidth]{ fig1.eps}
%\caption{(Color online) Schematics for high harmonics and attosecond pulse generations from the process of laser and plasma interaction. The plasma is produced by a prepulse (not showed). The following ultrastrong femtosecond IR pulse interacts with the plasma, and generate strong high harmonics. After filtering the high harmonics by filter, we can obtain a single attosecond pulse.}\label{fig:harmonic}
%\end{figure}

The majority of attosecond pulse research is currently focused on the creation of high-order harmonic generations (HHGs) from atomic and molecular gases \cite{Li17,Gaum17,Yue17,Han19, Guo19,wang20,Ozaki12} and solids \cite{Ghim12, Nava19, Ozaki11}, which is filtered to obtain attosecond pulses. According to the literature \cite{Cork07, Cork93, Kim08, Han19}, these methods of converting attosecond pulses are still at a low-level efficiency.
The use of plasma harmonics instead has the advantage of increasing the expected intensity of the attosecond pulses by orders of magnitude \cite{Lich96,Lich98,Thau10,Quer06,Bour83,Zhang2020,Zhong2020}. This feature will not only greatly ease the characterization of attosecond pulses, but also significantly expand the scope of their applications.
The HHGs emanating from the interaction of intense infrared laser pulses with overdense solid target surface plasma is an effective method to produce an attosecond pulse \cite{Lich96, Quer08}. 
Before the laser-plasma interaction (LPI) process, the plasma is prepared by a prepulse, i.e., heating pulse \cite{Konda21}, which is usually an ultrastrong femtosecond laser pulse. This prepulse hits the optically polished surface of the solid target and produces an overdense plasma that acts like a mirror which is called a plasma mirror. After that, an ultrastrong laser pulse is applied to the generated plasma. Because the plasma mirror reflects the high-intensity laser field, the nonlinear time response causes periodic time distortion of the reflected wave, which associates with the high-order harmonics in the frequency domain and the attosecond pulse in the time domain. After the generations of high-order harmonics, a filter is applied to obtain isolate-attosecond pulses.

Different from the HHGs from the gaseous targets which can be understood by the simple-man model, the HHGs from relativistic overdense plasma have different mechanisms, among which the two most common physical processes are the coherent wave emission (CWE) \cite{Quer06} and the relativistic oscillating mirror (ROM) \cite{Lich96,Beva06,Thau10}.
The CWE mechanism implies that the laser field pulls electrons from the plasma surface into the vacuum to absorb energy, and then push them back into the dense plasma. The fast electrons propagate through the dense part of the plasma, causing plasma oscillations to periodically occur in their wake. These collective electron oscillations radiate light at the different local plasma frequencies.% found in the inhomogeneous part of the plasma formed by the density gradient at the plasma vacuum interface.
 In the ROM mechanism where the incident laser is reflected by a high-speed oscillating mirror, the plasma surface electrons are accelerated by the laser field to the relativistic velocity with a high intensity. The laser field reflected by the dense plasma experiences a transient Doppler frequency upshift due to the movement of the plasma electrons. The Doppler effect produces a spectrum of laser harmonics because it occurs at periodic intervals with respect to the laser frequency.

%A. Tarasevitch {\it et al.} \cite{Alex07} and F. Dollar {\it et al.} \cite{Doll13} have demonstrated that the optimal scale length of the plasma gradient in the ROM mechanism and explore the effect of the scale length of plasma gradient on the high-order harmonics. The plasma gradient is a parameter of plasma that is controllable by experimentalists. In experiments, this parameter is adjusted by controlling the laser contrast through the plasma mirror (PM) systems.

For an obliquely incident $p$-polarized pulse, by adjusting its time delay and relative laser intensity to the heating pulse (prepulse), the electron density gradient scale length $L$ of the plasma can be adjusted in the experiments. The parameter $L$ depends strongly on the temporal contrast ratio of peak intensities between the driving and heating pulses \cite{Quer08}. Both the energy absorption of the plasma from the driving laser field and the subsequent emission of HHGs rely on the electron density gradient of the plasma formed by the heating pulse. In previous works, A. Tarasevitch {\it et al.} \cite{Alex07} and F. Dollar {\it et al.} \cite{Doll13} have explored the influence of $L$ on high-order harmonics, and the ideal scale length of the plasma gradient is found in the ROM mechanism is around $L$=0.2$\lambda$, where $\lambda$ is the wavelength of the driving pulse. 
This optimal gradient length may be extended to the peak intensity of an attosecond pulse, since it is filtered by a band-pass filter of HHGs. 
%The straightforward extension of this optimal gradient length to the attosecond peak intensity is that the optimal gradient length parameters are the same for the attosecond pulse since it is filtered by a band-pass filter of HHGs. 
%
Besides the peak intensity, the attosecond pulse is also characterized by the FWHM. In this work, we will show that the smallest FWHM has its own optimal parameter. %For different laser parameter, its optimal $L$ is different, but basically agrees with that of the pulse intensity at around $L$=0.2$\lambda$ at the higher laser intensities. 

The remainder of this work is structured as follows:
Section 2 presents the theoretical methods and ROM model for numerical simulations. The numerical results and discussion for the obtained HHGs and attosecond pulses are included in section 3. The section 4 contains the conclusions.

Unless specifically stated, atomic units (a.u.) $\hbar=m=e=1$ are used throughout.%, and work with the metric tensor $g^{\mu\nu}$=diag(1,-1,-1,-1).

\section{Theory}
The theoretical basis for LPI to produce HHGs is to solve Maxwell's equations for the fields and relativistic Newton equation for macroparticles. The HHGs spectrum is obtained by calculating the dipole momentum of the electrons accompanied by a Fourier transform into the frequency domain. The common numerical method is to use the particle-in-cell (PIC) algorithm to divide the space into cells where the particles equations are solved \cite{Quer06,Lich96,Lich98}. Motivated by the prospects offered by the rapidly evolving laser technology, we have adopted the one-dimensional PIC code \cite{Lich96} in the  current work .

{\it Basic equations} - The plasma is subject to the Maxwell's equations of the electromagnetic field and the Newton equations of the macro electron in each cell. %
The Maxwell's equations read
\begin{equation}
\begin{split}
  \nabla \cdot \mathbf{B} &= 0, \\
  \nabla \times \mathbf{B} &= \mu_0\varepsilon_0\frac{\partial E}{\partial t}+\mu_0\mathbf{J}, \\
  \nabla \cdot \mathbf{E} &= \frac{\rho}{\varepsilon_0}, \\
  \nabla \times \mathbf{E} &= -\frac{\partial B}{\partial t}.
\end{split}
\end{equation}
The velocity, acceleration and position of the charged particles are solved by the relativistic Newton equations,
\begin{equation}
\begin{split}
  \frac{d \mathbf{p}}{dt} &= q(\mathbf{E}+\mathbf{v\times B}), \\
  \mathbf{p} &= m\gamma \mathbf{v}, \\
  \mathbf{v} &= \frac{dx}{dt}, \\
  \gamma &= \sqrt{1+\left(\frac{\mathbf{p}}{mc}\right)^2},
\end{split}
\end{equation}
where $m$ is the mass of electron.

To solve the oblique incidence problem, a Lorentz transformation is performed from the laboratory frame to a frame that moves in the direction parallel to the surface, In such a transformed frame, the laser pulse incident normally.
In PIC simulation, this procedure is iterated leading to the self-consistent evolution of plasma and field.
For the kinetic simulations, we use a 1D3V relativistic electromagnetic PIC Code (LPIC++) which was proposed by R. Lichters {\it et al.} \cite{Lich96} in 1996. This PIC code considers only one spatial dimension, but particle velocity in three spatial directions. The merits of this method have been realized in a variety of applications by different groups \cite{Lich96, Lich98, Quer06, Thau10}.

{\it ROM model} - As mentioned above, high-order harmonics are produced by the Doppler effect. To simplify things, we'll only discuss the normal incident here. Based on the fundamental electrodynamics theory \cite{Thau10},

\begin{equation}
\omega^{\prime}=\sqrt{\frac{1+\beta}{1-\beta}}\omega,
\end{equation}
where $\omega$ is the laser frequency in the reference frame, $\omega ^{\prime}$ is the frequency observed by plasma mirror, and $\beta=\frac{v}{c}$, in which $v$ is the speed of plasma. When the laser is reflected from the oscillating mirror and caught by the sensor, the Doppler effect happens again,
\begin{equation}
\omega ^{\prime\prime}=\sqrt{\frac{1+\beta}{1-\beta}}\omega ^{\prime}=\frac{1+\beta}{1-\beta}\omega.
\end{equation}
It's the traditional Doppler shift in the oscillating mirror, $\omega ^{\prime\prime}\rightarrow 4\gamma ^{2} $ when $v\rightarrow c$. Previous researches \cite{Lich96, Lich98, Quer06, Thau10} have found that HHGs from ROM are emitted simultaneously when $\gamma$ reaches its maximum, therefore, the maximum value of $v$ determines the highest order that ROM can reach.

\section{Results and discussion}
In the present work, we restrict the numerical simulations in one-dimensional geometry for a plane of plasma layer with sharp surface.
Since the plasma ions move slowly on the time scale of the short pulses we applied, it is reasonable to assume that the ions remain fixed during the laser-plasma interactions. We make the further approximation that the plasma extends linearly towards the vacuum.
%plasma extends exponentially towards the vacuum.

%The selection rules is the relationship between the characteristics of the high-order harmonics and the polarizations incident laser pulses. Theoretically, these rules can be explained by the motion of the relativistic electrons which follow a ``figure-of-8'' path [{\color{red} add citations}].

\begin{figure}[htb]
\includegraphics[width=.91\columnwidth]{ 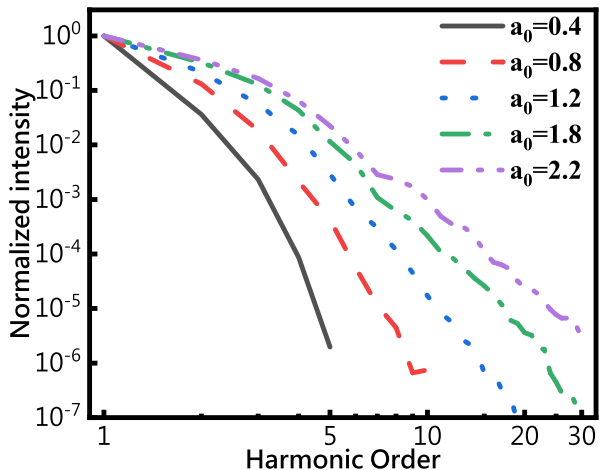}
\caption{(Color online) The profiles of normalized HHGs from plasma by different laser intensities. The laser parameters are as follow: 800 nm wavelength infrared laser pulse with central frequency $\hbar \omega$=1.55 eV, and the $p$-polarized pulse duration is 20 fs with sinusoidal profile envelop. The incident angle $\alpha$=45$^{\circ}$, $n_e/n_c$=15.}\label{fig:frequency}
\end{figure}

It is interesting to investigate the effects of laser intensity on the HHGs first.
As revealed by Quere {\it et al.} \cite{Quer06} in their previous work, in the case of $a_0<$1 ($a_0=$1 corresponding to the laser intensity $I\lambda^2=1.37\times10^{18}$ W$\cdot\mu$m$^2$/cm$^2$ for 800 nm NIR laser), the HHGs mechanism is dominated by the CWE mechanism; while for $a_0>$1, both CWE and ROM exist simultaneously, but they active in different density regions of plasma \cite{Quer08} and their relative weight is determined by laser intensity.
Therefore, we explore the relationship between the conversion efficiency and intensity of the incident laser. Fig.~\ref{fig:frequency} demonstrates that a higher laser intensity leads to greater efficiency of the high-order harmonics. We display only the conversion yield that higher than 10$^{-7}$. For $a_0$=0.4, where the CWE mechanism is dominant, we can only observe a few orders of harmonics at very low efficiency. When the laser intensity is increased to $a_0$=1.2, the harmonics with order higher than 10 are observed. 
This is due to the fact that the ROM radiation from the Doppler effect emerges with considerable contributions.
When the laser intensity becomes higher, such as $a_0$=1.8 and 2.2, one expects higher-order harmonics and better conversion yield.

 More details about the HHGs in each order are shown in Fig.~\ref{fig:hhgvsintensity}.
The peak intensities of HHGs at selected values of harmonic order are presented as functions of the incident laser intensity. When the laser intensity is increased, the harmonic signal increases and the low-order harmonics (e.g., the 2nd, 3rd, and 5th orders) become saturated. The higher orders of harmonics (e.g., the 10th, 20th, 30th, and 40th orders) are also developed. Therefore, an appropriate choice of laser intensity can greatly boost the conversion efficiency of higher-order harmonics.

\begin{figure}[htb]
\includegraphics[width=.91\columnwidth]{ 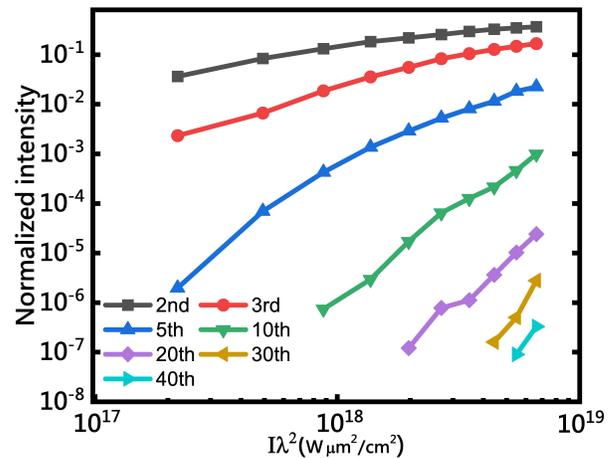}
\caption{ Normalized HHGs' intensities of some typical orders from the LPI process with different laser intensities. The laser and plasma parameters are: 800 nm wavelength infrared laser pulse with frequency $\hbar \omega$=1.55 eV, and the $p$-polarized pulse duration is 20 fs with sinusoidal profile envelop. The incident angle $\alpha$=45$^{\circ}$, plasma density $n_e/n_c$=15.}\label{fig:hhgvsintensityy}
\end{figure}

%{\it Size length of optimal plasma gradient in the ROM mechanism}
%This section explores the effect of plasma size length, i.e. gradient steeple, on high harmonic intensity and a second pulse strength. In experiments, plasma mirror (PM) systems are usually used to improve laser contrast control.
%To explore the effect of $L$ in ROM on the efficiency of high-second harmonic conversion, we selected the following different dimension lengths. Figure 3.5 shows the strength of HHG under different Ls. The results show that HHG is the largest in L-0.2. Figure 3.6 selects the strength of the 10th, 20th and 25th-order harmonics with L. The results also show that, with the increase of L, the HHG strength increases rapidly, then becomes saturated. The optimal value is in $L$=0.2$\lambda$.

\begin{figure}[htb]
\includegraphics[width=1\columnwidth]{ 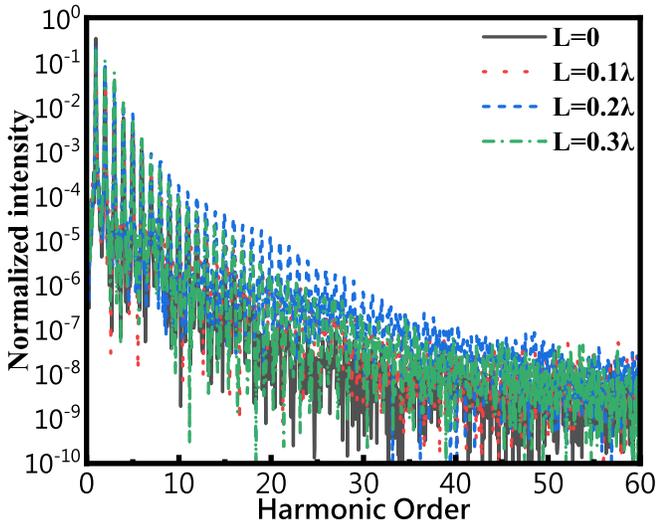}
\caption{ Power spectrum of HHGs with different length of plasma density gradient. The laser parameters are: $\hbar \omega$=1.55 eV, FWHM is 20 fs, $a_0=2$, linearly $p$-polarized, sinusoidal profile envelop. Incident angle $\alpha$=45$^{\circ}$, $n_e/n_c$=15.
}\label{fig:hhglog}
\end{figure}

In the following discussion, we focus on the 10th and higher-order harmonics and investigate the optimal laser parameters and plasma characters in the conversion of HHGs and attosecond pulses.
% A. Tarasevitch {\it et al.} \cite{Alex07} and F. Dollar {\it et al.} \cite{Doll13} have explored the optimal scale length of the plasma gradient in the ROM mechanism and  demonstrated the effect of the scale length on the high-order harmonics. The plasma gradient is a parameter of plasma that is controllable by experimentalists. In experiments, this parameter is adjusted by controlling the laser contrast through the plasma mirror systems.
%
In laser-plasma experiments, the plasma gradient is an important parameter that can be adjusted by tuning the laser contrast through the plasma mirror systems. In the recent works of A. Tarasevitch {\it et al.} \cite{Alex07} and F. Dollar {\it et al.} \cite{Doll13}, these authors have successfully explored the optimal scale length of the plasma gradient in the ROM mechanism and demonstrated the effect of the scale length on the high-order harmonics.
%
%have successfully demonstrated the effects of scale length (i.e. L) for plasma gradient on the high-order harmonics and obtained the optimal value of L in the ROM mechanism at about 0.2 lambda (?). 
Therefore, it is of great interest to explore the effect of scale length on the conversion efficiency of high-order harmonics and isolate-attosecond pulses. 
%
%To explore the effect of $L$ in ROM model on the conversion efficiency of high-order harmonics, we consider different scale lengths for the plasma gradient.
Fig.~\ref{fig:hhglog} shows the intensity of high-order harmonics with different values of $L$ at 0, 0.1$\lambda$, 0.2$\lambda$, and 0.3$\lambda$. 
The situation with $L$ = 0 which corresponds to no density gradient in the plasma has the lowest conversion efficiency.
%
%The $L$ = 0 curve has the lowest conversion efficiency of them. 
%The $L$ = 0.2$\lambda$ and 0.3$\lambda$ curves have a better yield in the higher orders of harmonics. 
Our simulations with L=0.2$\lambda$ and 0.3$\lambda$ have better yields in both low and high orders of harmonics.
%{\it Effect of laser intensity on HHG process}
%This section explores the effects of laser intensity. At $a_0<$1, the high harmonic generation mechanism is the CWE mechanism. At $a_0>$1, both CWE and ROM exist at the same time, but are active in different density regions.
%Therefore, the relationship between the intensity of incident laser and conversion efficiency under the ROM mechanism is explored. Figure 3.5 shows that the stronger the laser intensity in this intensity range, the higher the efficiency of high-level harmonic conversion. Figure 3.6 selects the strength of several harmonics with the change of laser intensity, the visible harmonic signal increases with the increase of laser intensity, and then becomes saturated. Therefore, choosing the appropriate incident laser intensity is a way to improve the efficiency of harmonic conversion.

\begin{figure}[htb]
\includegraphics[width=.91\columnwidth]{ 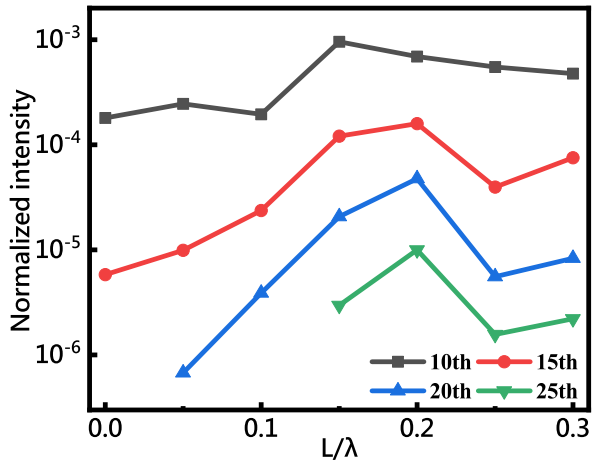}
\caption{ Normalized intensity of HHGs with different plasma electron density gradient length $L$.The laser and plasma parameters are the same as the Fig.~\ref{fig:hhglog} unless stated in graphs. % The laser parameters are: $\hbar \omega$=1.55 eV, FWHM is 20 fs, $a_0=2$, linearly polarized, Sinusoidal profile envelop. Incident angle $\alpha$=45$^{\circ}$, $n_e/n_c$=15.
 }\label{fig:hhgvsorder}
\end{figure}

%Then we'll get into the specific intensities of various orders of HHGs.%, and choose the 10th, 15th, 20th, and 25th harmonics. %we found that the intensity of HHG increases rapidly, then becomes stable with L. 
%
In Fig.~\ref{fig:hhgvsorder}, we present the normalized yield of the 10th, 15th, 20th, and 25th orders of harmonics versus the plasma gradient $L$. Except that the 10th order harmonic has a maximum at $L$=0.15$\lambda$, the other harmonics have their maxima at $L$=0.2$\lambda$. This is in agreement with the previous simulations \cite{Alex07}. The exception for the 10th order can be attributed to the contribution from CWE emission which is most significant in lower orders of harmonics \cite{Quer08,Quer06}. Regardless this exception, the optimal value of gradient length for the plasma density is about $L$ = 0.2$\lambda$.

HHGs offer the enticing prospect of synthesizing EUV/soft x-ray pulses of attosecond duration.
We shall now concentrate on the  creation of isolated attosecond pulses from the LPI.
Attosecond pulses are created by slicing a portion of the harmonic spectrum using a band-pass filter between the 20th and 60th orders of harmonics.
Two typical characteristics for an attosecond pulse are the FWHM and peak intensity. 
In our following work, we will investigate the optimal scale length of the plasma gradient for both of these two characteristics.

\begin{figure}[htbp]
\centering
\includegraphics[width=.4955\columnwidth]{ 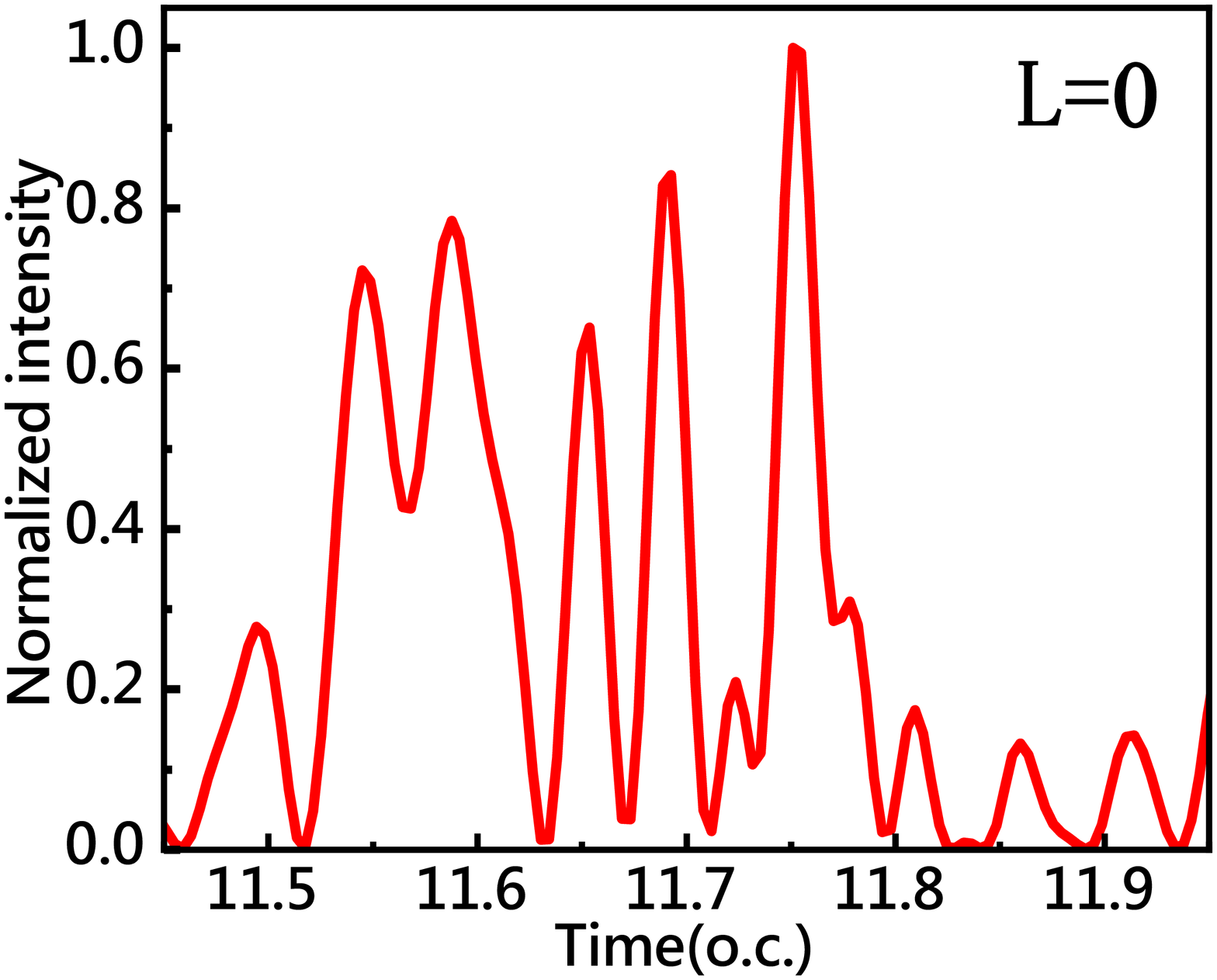}
\includegraphics[width=.4955\columnwidth]{ 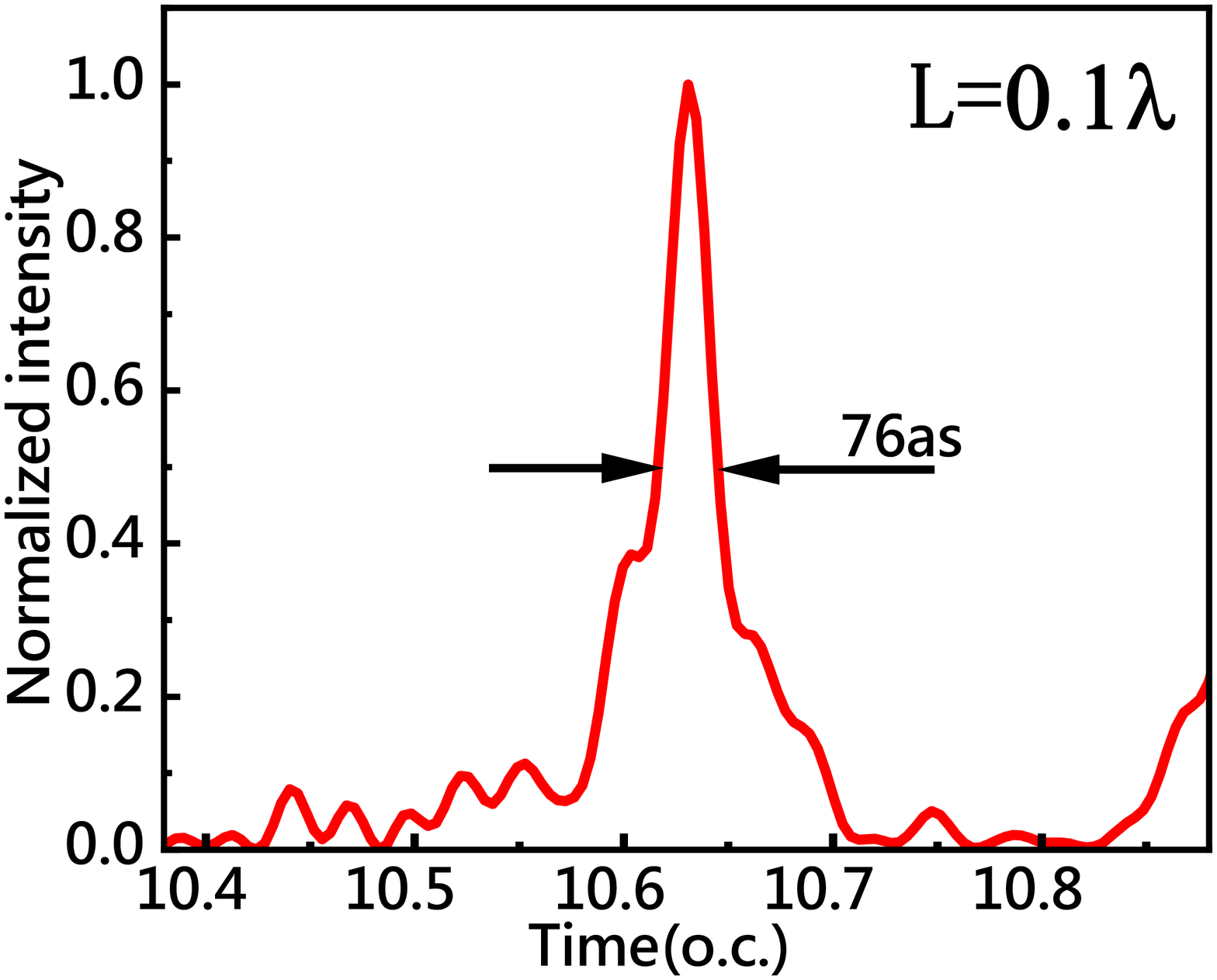}
\includegraphics[width=.4955\columnwidth]{ 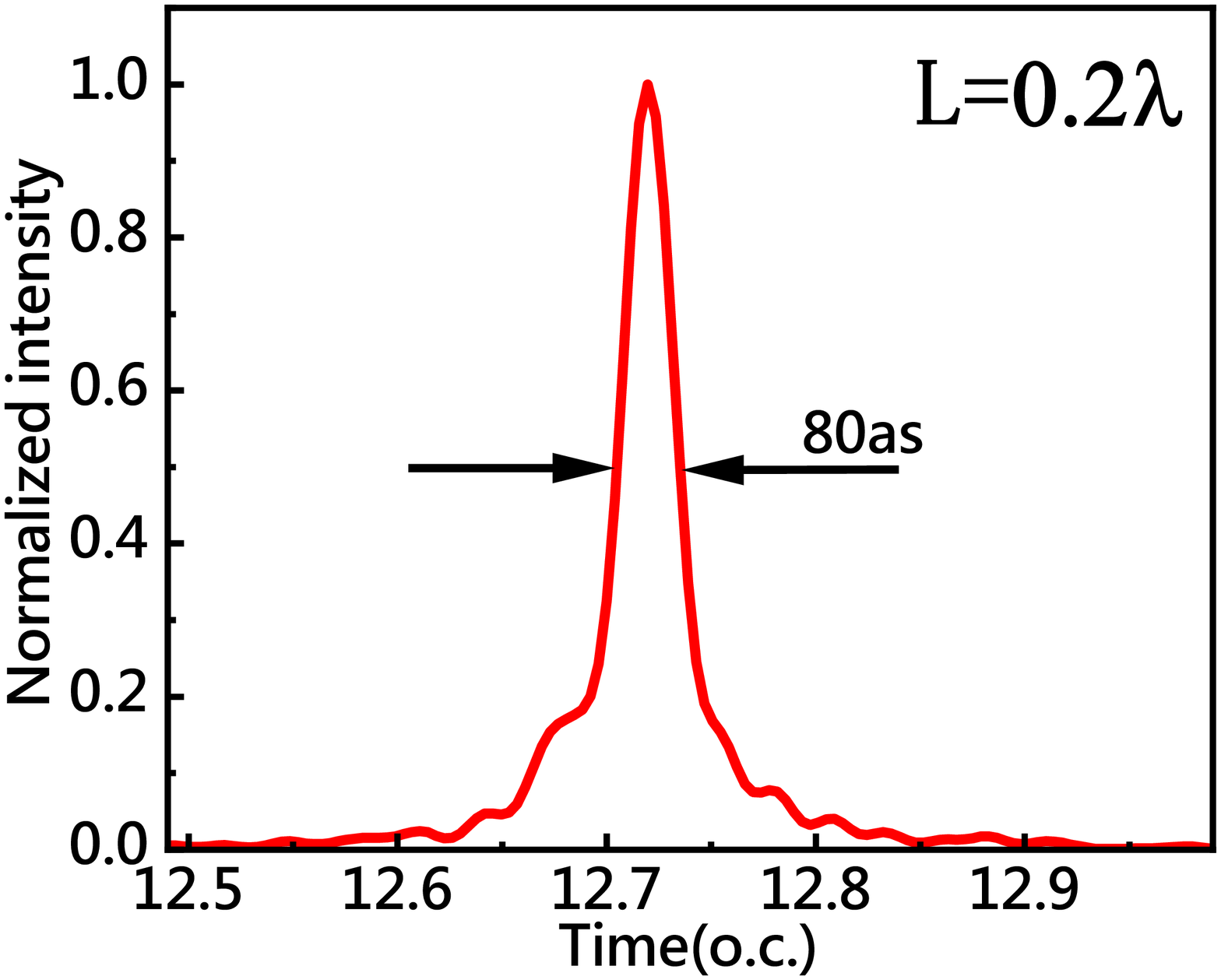}
\includegraphics[width=.4955\columnwidth]{ 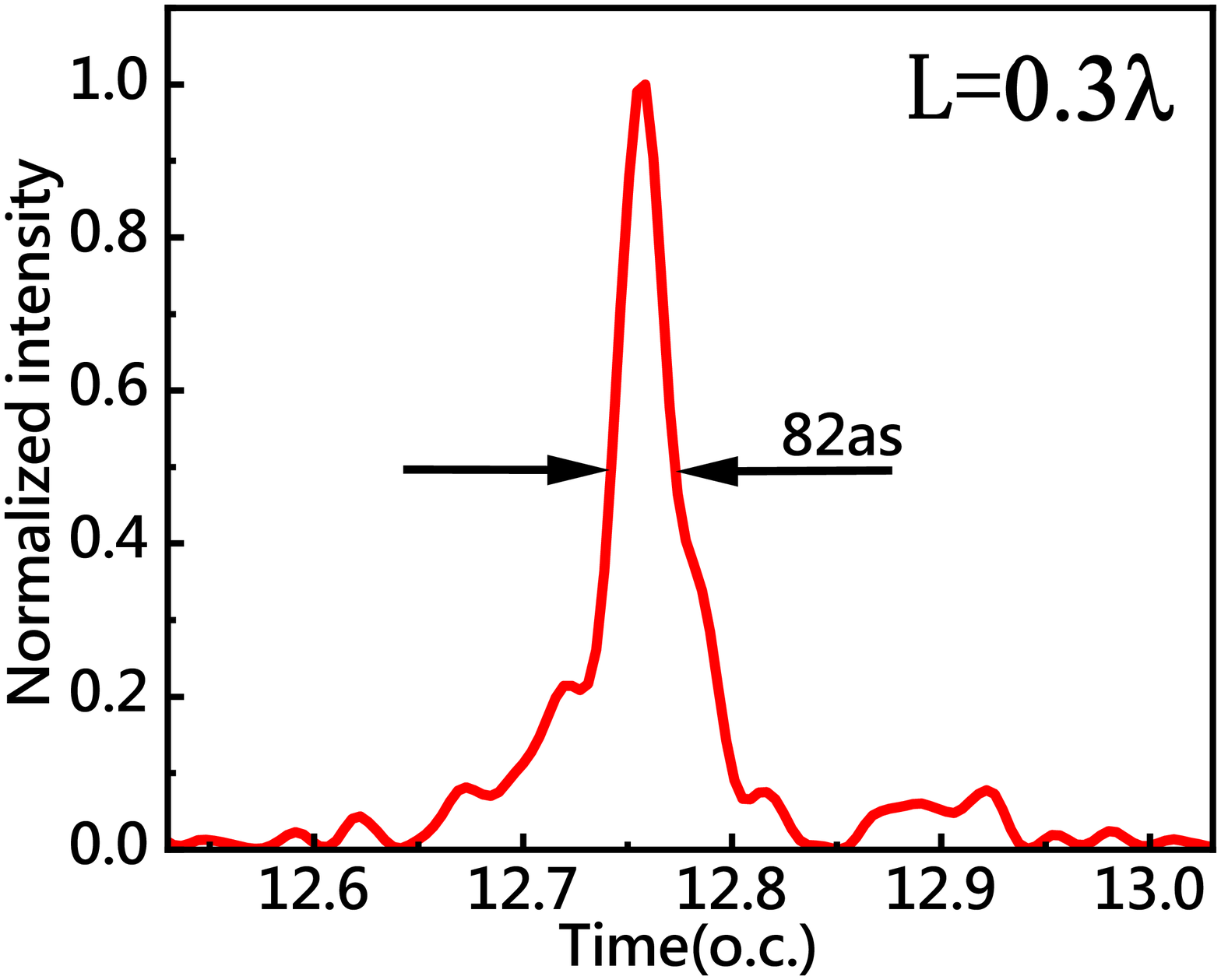}
\caption{Attosecond pulses gained by band-pass filtering of HHGs between the 20th and 60th orders when $a_{0}$=2. The laser and plasma parameters are the same as the Fig.~\ref{fig:hhglog} unless stated in graphs. %The laser parameters are: 800 nm wavelength near infrared laser pulse with frequency $\hbar \omega$=1.55 eV, and the $p$-polarized pulse duration is 20 fs with sinusoidal profile envelop. The incident angle $\alpha$=45$^{\circ}$, $n_e/n_c$=15.
}\label{fig:asp}
\end{figure}

\begin{figure}[htbp]
\includegraphics[width=1.\columnwidth]{ 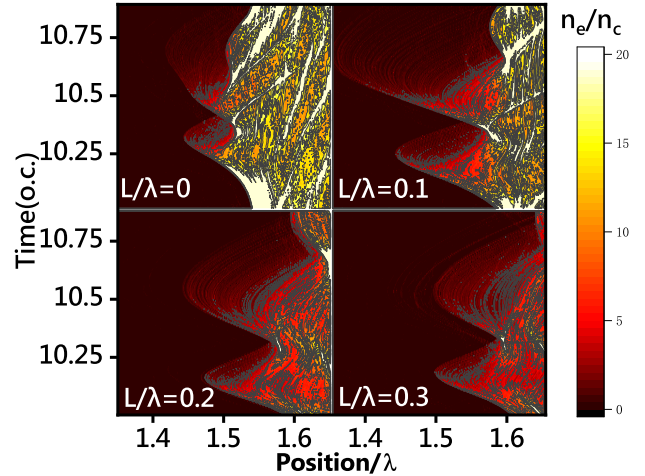}
\caption{ The space-time evolution of the plasma electron density with different gradient length $L$. The laser and plasma parameters are the same as the Fig.~\ref{fig:hhglog} unless stated in graphs. %The incident angles are 45$^{\circ}$ for all graphs.
}
\label{fig:spacetime}
\end{figure}

In Fig.~\ref{fig:asp}, we present the normalized attosecond pulses with laser parameters of $a_0$=2, FWHM=20 fs, and the plasma parameter $n_e/n_c$=15. The incident angle is 45$^{\circ}$, and the gradient lengths selected are 0, 0.1$\lambda$, 0.2$\lambda$, and 0.3$\lambda$. It is hard to recognize a well-isolated attosecond pulse when $L$=0. In the remaining cases, we obtain attosecond pulses with FWHMs of 76, 80, and 82 as for $L$ = 0.1$\lambda$, 0.2$\lambda$, and 0.3$\lambda$, respectively. From the discussion of HHGs mentioned above, it is known that the best harmony between plasma restoring force and the plasma oscillation is achieved at $L$ = 0.2$\lambda$, which gives the highest yield of HHGs. In considering the FWHM, a stronger restoring force, i.e., a smaller value of $L$ is required to get shorter FWHM. %{\color {red} (REPEAT ?)} This means that one needs a smaller value of L to get shorter FWHM. 
However, when $L$ approaches 0, the production of HHGs and attosecond pulses becomes weaker due to the ultrastrong restoring force. As a results, it is hard to acquire a well-isolated attosecond pulse. 
Therefore, one expects that the optimal density gradient length should be close to but not exactly equal to zero.
%
%As a result, the optimal density gradient length should be near to 0 but cannot be exactly.
%
The attosecond pulse can be optimized by adjusting the plasma gradient length $L$ as well. Neither a too long nor a too short gradient length is beneficial to the high-harmonics generations. This is demonstrated with the help of a space-time diagram of electron density shown in Fig.~\ref{fig:spacetime}. For $L$<0.2$\lambda$, plasma's oscillation is not significant because of high restoring force. The plasma oscillation is improved with the increase of $L$, which corresponds to the decrease of restoring force. When $L$=0.3$\lambda$, oscillation is stable but has relatively lower density and amplitude. The combination of these effects leads to an optimal value at $L$=0.2$\lambda$.

\begin{figure}[htbp]
%\subfigure[:FWHM]{
\includegraphics[width=.5\columnwidth]{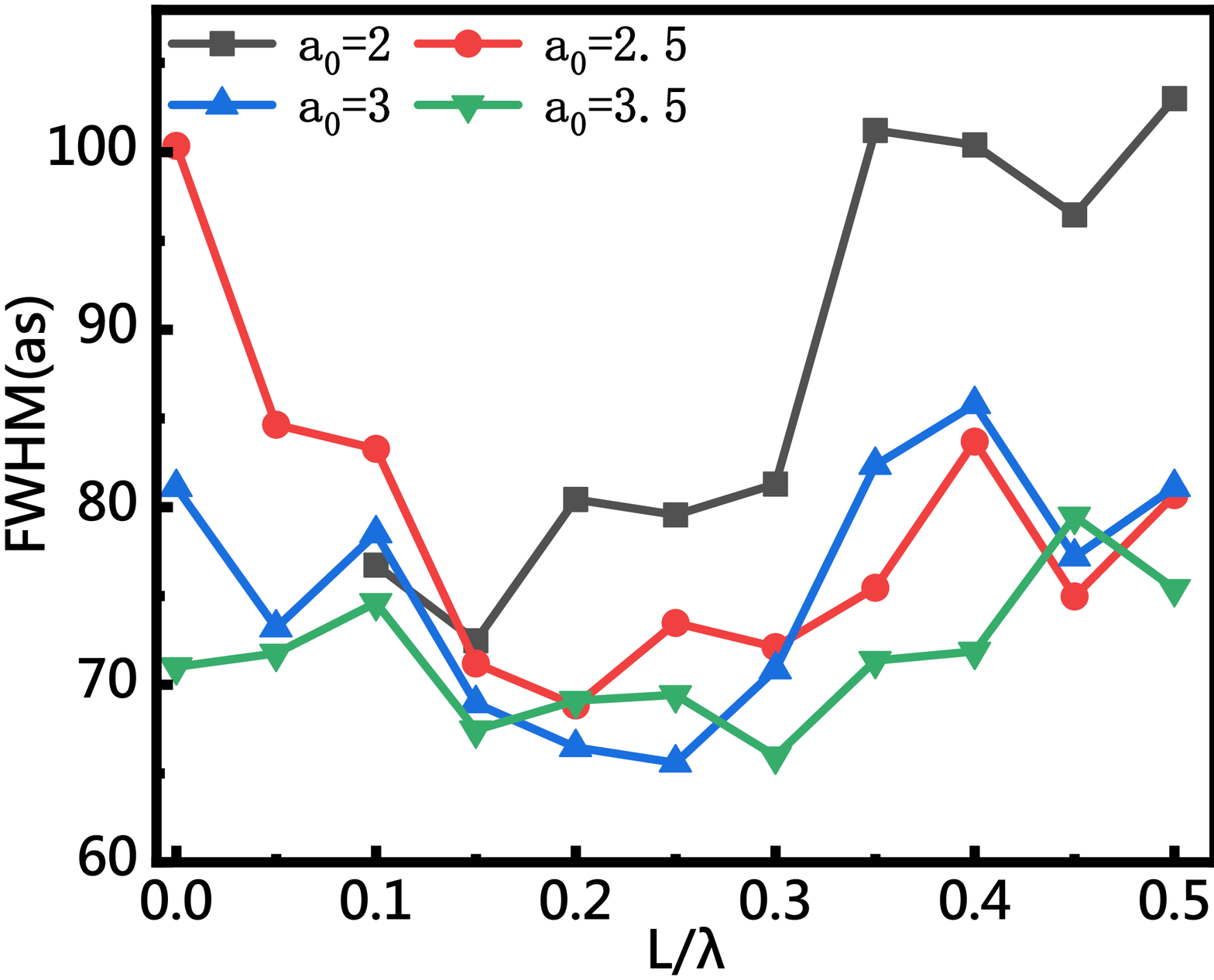}%}
\includegraphics[width=.5\columnwidth]{ 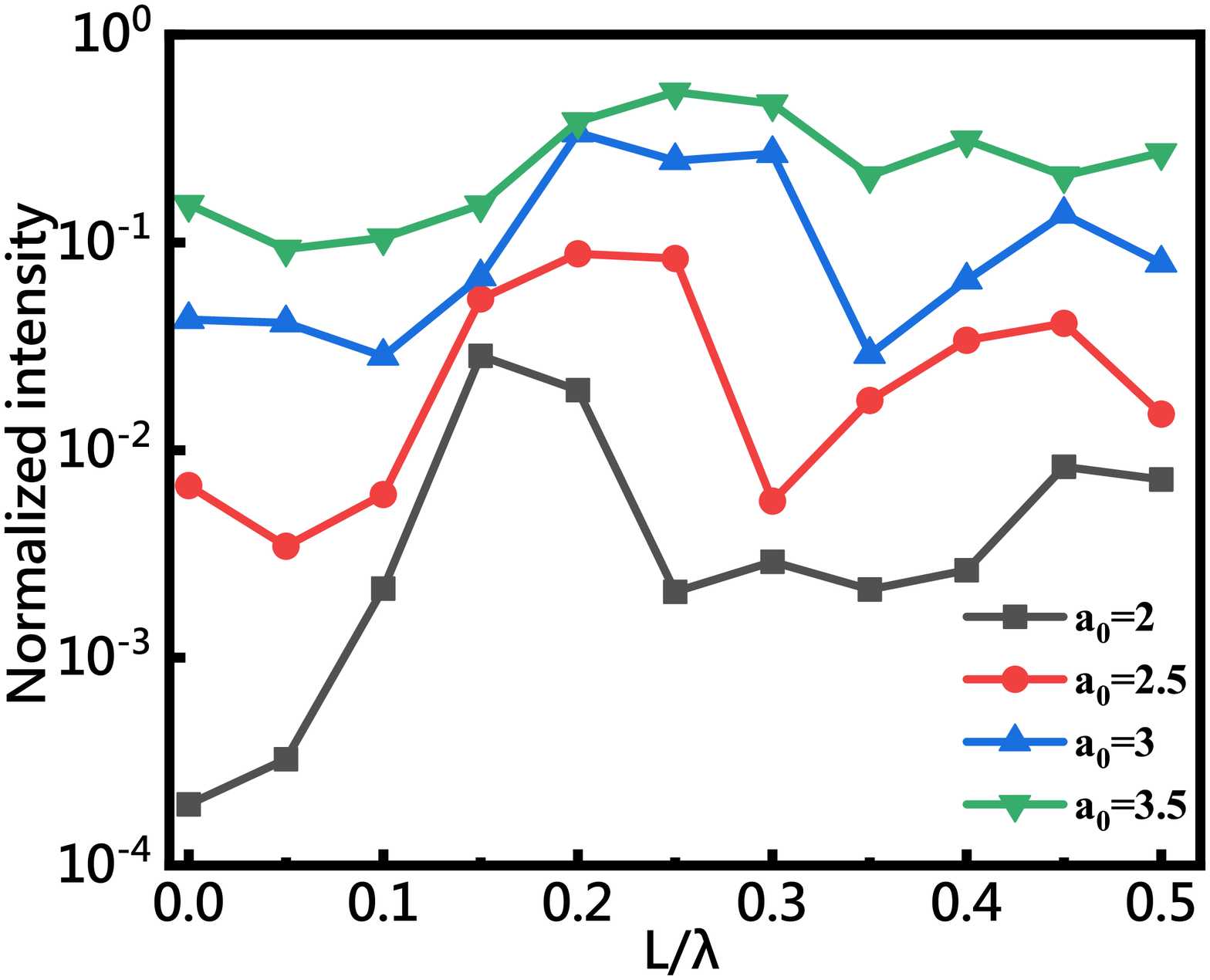} %}%\subfigure[:Peak Intensity]{
\caption{ Scaled plasma gradient length ($L/\lambda$) dependencies of the FWHM (left panel) and peak intensity (right panel) of filtered attosecond pulse with different driving laser intensities when frequency $\hbar \omega$=1.55 eV, $p$-polarized pulse; the duration is 20 fs with sinusoidal profile envelop. The incident angle $\alpha$=45$^{\circ}$, $n_e/n_c$=15.
}\label{fig:aspintensity}
\end{figure}

In Fig.~\ref{fig:aspintensity}, we present the FWHM (left) and peak intensity (right) of attosecond pulses with different laser parameters and gradient lengths. 
%Due to the strong restoring force and low yield of HHGs, the isolation of attosecond pulse is bad in small gradient length and gives relatively large FWHM.
Due to the strong restoring force of plasma and low yield of HHGs, the isolation of attosecond pulse is poor at small gradient lengths (e.g., $L$<0.1$\lambda$) which results in a relatively large FWHM.
For the driving laser intensity $a_0$=2, the shortest FWHM  is obtained at $L$=0.15$\lambda$. As the laser intensity increases, the FWHM decreases to as short as only 66 as at a larger gradient length between $L$=0.2$\lambda$ and 0.3$\lambda$.
From the peak intensity curves displayed in the right panel, it is found that the highest yields do not exactly agree with the yield of HHGs. For the lower intensity $a_0$=2, the highest yield is obtained at $L$=0.15$\lambda$. As the laser intensity increases, the highest peak intensity increases and they are generally located at about $L$=0.2$\lambda$. In all curves, there are minima when $L$ is lager than 0.2$\lambda$ if the gradient length surpasses the harmony gradient length between restoring force and plasma oscillating, where the chaos between them is boosted. However, for $a_0$=3.5, the curve is less dependent on gradient length for L>0.2$\lambda$. This is due to the fact that the restoring force in this scenario is comparably smaller than the interaction caused by the prevailing, powerful laser field.
%
%As discussed above in the Fig.~\ref{fig:hhgvsorder}, we know that the stronger laser pulse can boost the higher order HHGs much better that the lower order HHGs. In the generation of attosecond pulses, we only keep the higher order HHGs, and filtered the lower orders, therefore, the best gradient length for HHGs do not exactly agree with that of isolate attosecond pulses. The stronger pulses generate higher yield of higher order harmonics.
%

 \section{Conclusions and final remarks}
In this work, we have investigated the optimal plasma density gradient length for the HHGs and isolated-attosecond pulse creation in the laser-plasma interactions. The optimal gradient length for HHGs is confirmed with previous value in the literature. The pulse duration and peak intensity of the attosecond pulses depend on both the driving laser parameters and the scaled plasma gradient length. It has been shown that the best gradient length for producing isolated-attosecond pulse with the shortest pulse duration is different from the one with the maximum peak intensity, however, in high laser intensities they are both located at about $L$=0.2$\lambda$. It is hoped that the present study could provide useful information for increasing the productions of HHGs and the generation of isolated-attosecond pulses in further theoretical developments and experimental measurements.

%The application of LPI to produce HHG and attosecond pulses still has a lot of outstanding issues. Future studies may focus on: (1) using harmonics as a diagnostic tool for LPI; (2) controlling harmonics, achieving higher conversion efficiency, harmonic power, spectral width.

\section{Funding}
This work is supported by the National Natural Science Foundation of China under Grant Numbers 91850114, 11774131 and 11974007.

\section{Acknowledgments}
L. Cheng and A. Liu thank L.Y. Peng for helpful advice and suggestion. Part of the numerical simulation was done on the high-performance computing cluster Tiger@IAMP in Jilin University.

\bibliography{ref} % Produces the bibliography via BibTeX.
%\begin{thebibliography}{99}

%\end{thebibliography}

\clearpage
%  \section*{FIGURE CAPTIONS}
%

%\end{CJK*}  %% end the Chinese environment

\end{document}